\documentclass[12pt]{article}\textheight 24.5cm \textwidth 17cm\voffset=-1.in\hoffset= - 1.in

\usepackage[utf8]{inputenc}
\usepackage[T2A]{fontenc}
\usepackage{amsmath}
\usepackage{slashed}
\usepackage{graphicx}
\usepackage{amsfonts}
\usepackage{amssymb}
\usepackage{psfrag}
\usepackage{hhline}
\usepackage{cite}
\usepackage{upgreek}
\usepackage{epsfig,amsfonts}
\usepackage{feynmp}
\author{ Lev Spodyneiko$^{1,2}$\footnote{E-mail:  lionspo@itp.ac.ru}
\vspace*{10pt}\\[\medskipamount]
$^1$~\parbox[t]{0.88\textwidth}{\normalsize\it\raggedright
L.D.Landau Institute for Theoretical Physics,
142432 Chernogolovka, Russia}
\vspace*{10pt}\\[\medskipamount]
$^2$~\parbox[t]{0.88\textwidth}{\normalsize\it\raggedright
Moscow Institute of Physics and Technology, 141700 Dolgoprudny, Russia}
}
\date{}
\title{\bf Minimal Liouville Gravity on the Torus via Matrix Models}

\begin{document}

\maketitle

\begin{abstract}
In this paper we use recent results on resonance relations between the matrix models and the minimal Liouville gravity to compute the torus correlation numbers in $(3,p)$ minimal Liouville gravity. Namely, we calculate the torus generating partition function of the $(3,p)$ matrix models and use it to obtain the one- and two-point correlation numbers in the minimal Liouville gravity.

\end{abstract}

\section{Introduction}

In present paper we study  the two-dimensional quantum gravity. Continuous approach to the quantum gravity is based on the functional integration over the two-dimensional geometries
\begin{equation}
 Z \sim \sum_{h=0}^{\infty} \int \mathcal D g \,\mathcal D X\, e^{-S},
 \end{equation}
where $g$ is the metrics, $X$ is a set of matter fields and the summation goes over the different topologies labeled by the genus $h$.

 It was shown in \cite{Polyakov:1981rd} that after choosing conformal gauge for the metrics, theory becomes a tensor product of two almost independent parts: the initial matter part and the Liouville theory of scalar field. This approach is called Liouville gravity. See also \cite{KPZ}.
The computation of the correlation numbers in the Liouville gravity is a complicated problem connected with the  integration over the moduli space. Recently some progress in it was made in \cite{Zamolodchikov:2003yb,Belavin:2006ex}, where it was noticed that the problem of the computation of the correlation numbers simplifies in the case of so called Minimal Liouville gravity (MLG), whose matter sector being Minimal Model of CFT  \cite{Belavin:1984vu}.  With use of higher equations of motion of the minimal Liouville gravity \cite{Zamolodchikov:2003yb} four-point correlation numbers for the genus zero surfaces were determined by Belavin and Zamolodchikov in \cite{Belavin:2006ex}. We will call the $(q,p)$ MLG a LG model with matter fields from a $(q,p)$ minimal model of CFT with central charge $c = 1- 6\frac {(p-q)^2}{pq}$. One can define a deformed partition function of the MLG as
$$Z[\lambda] = \left \langle\exp \sum_{m,n} \lambda_{m,n} O_{m,n} \right\rangle ,$$
where $\lambda_{m,n}$ are coupling constants and $O_{m,n}$ are local observables introduced later in the section 3. This is a function  of particular interest since correlation numbers of MLG can be obtain from it by differentiation with respect to the couplings $\lambda_{m,n}$.

Another approach to the summation over fluctuating two-dimensional surfaces was proposed in the late '80s\cite{Kazakov:1985ea,Gross:1989aw,Kazakov:1986hu,Kazakov:1989bc,Staudacher:1989fy,Brezin:1990rb,Douglas:1989ve}. It was based on the idea of approximation of a continuum two-dimensional surface with its discrete triangulation and the replacement of the functional integral over continuum surfaces with finite-dimensional ones. Since it was technically equivalent to the integration over matrices this approach is commonly called the Matrix Model of two-dimensional gravity.  The continuous surfaces are retained in a certain limit in which surface with large number of vertices dominate. This limit can be achieved by taking the size of the matrix to infinity along with the adjustment of the parameters of the action. This procedure of tuning the coupling to get the continuous limit in which initial "lattice" disappears is similar to the second order transition in the condensed matter physics. Moreover, just as in case of the phase transition there are exist multi-critical point in matrix models.  We will label model containing $q-1$ matrices at its $p$-critical point as the $(q,p)$ matrix model of two-dimensional gravity. We will be interested in the partitions function of the matrix models $Z[\tau]$ in the vicinity of its critical point. The deviations from the critical point are parameterised by times $\tau_{m,n}$.

Douglas had shown in \cite{Douglas:1989dd} that the computation of the partition function $Z[\tau]$ of the matrix models could be reformulated in the terms of so called "string equation".  It was also noticed, that times $\tau_{m,n}$ have that same scaling dimension as the couplings $\lambda_{m,n}$ in the Liouville gravity do. Therefore, it seems natural to expect that
the matrix models  have the same correlation numbers, critical exponent etc. as the Liouville gravity does. However, it was shown that the correlation numbers of the matrix models don't satisfy the fusion rules of CFT. For example, the conformal symmetry forces all one-point correlation functions on the sphere in the MLG to be zero, but  they  don't vanish in the matrix models. Moore et al. \cite{Moore:1991ir} proposed a solution to this problem. The main idea was that the times $\tau_{m,n}$ and the coupling $\lambda_{m,n}$ are related in a non-linear fashion
\begin{equation}\label{res rel intro}
\tau_{m,n} = \lambda_{m,n} + \sum_{m_1,m_2,n_1,n_2} C^{m_1n_1,m_2n_2}_{m,n}\lambda_{m_1n_1}\lambda_{m_2n_2}+\dots.
\end{equation}
Once one computed correlation numbers of the matrix models, requirement for the correlation numbers in the MLG to satisfy the fusion rules gives equations for the coefficients in the resonance relation (\ref{res rel intro}), from which they could be determined. Therefore, this provides an indirect method of computation of the correlation numbers in the MLG. Moore et el. found some first of this coefficient in order to saturate one- and two-point numbers fusion rules. Their work were continued in \cite{Belavin:2008kv,Belavin:2013nba,Tarnopolsky:2009ec}, where mentioned technique were used to compute three-, four- and five-point correlation numbers. Also in \cite{Belavin:2013nba} there was found an interesting connection of the matrix models with the Frobenius manifolds.

All the mentioned papers concerned amplitudes on the sphere. Extension of these results to the torus was done for the $(2,p)$ models in \cite{Belavin:2010pj, Belavin:2010bs}. The logic behind these papers is the following. The resonance relations can be computed by use of the explicit form of the correlation functions of the matrix models on the sphere and use of the fusion rules of the MLG. Because of their local nature the resonance relations must be independent of the genus. Therefore, they also give the correspondence between the torus correlation function of the matrix models and the MLG. So by this indirect procedure the torus correlation function in the MLG can be determined. Note, that there is no direct computation of the torus correlation functions in the MLG. Therefore, this methods gives a proposal for them.  For the case of $(2,p)$ it was done in \cite{Belavin:2010pj} and their results were confirmed by the computer computations carried out in \cite{Belavin:2010sr}.

 The aim of the present work is to extend these results to the case of the $(3,p)$ models: to find the genus one partition function in the $(3,p)$ matrix models and to calculate the torus one- and two-point correlation numbers in the $(3,p)$ MLG with  use of the exact form of the resonance relation, which was carried out in \cite{Belavin:2013nba}.

The outline of the paper is the following. In section 2 we review the matrix models and present the results for the $(3,p)$ matrix models partition function. We provide basic information about MLG in section 3. In section 4 we calculate one- and two-point correlation numbers of the matrix models and the MLG. We left the technical details of the partition function in the appendix A. Determination of the correlation number and the partition function is the main result of our paper.

\section{Matrix Models}

The matrix models approach appeared as a technical tool to calculate integrals over the discrete two-dimensional surfaces. The model with $q-1$ $N\times N$ hermitian matrices $M^{(\alpha)}$ has the partition function defined as
\begin{equation}
 Z = \int \prod_{\alpha=1}^{q-1}dM^{(\alpha)} e^{\text{tr\,}\left(-\sum_{\alpha=1}^{q-1}V_\alpha(M^{(\alpha)}) + \sum_{\alpha=1}^{q-2}c_\alpha M^{(\alpha)}{M^{(\alpha+1)}}\right)}.
 \end{equation}
where $V_\alpha(x)$ are polynomial potentials, $c_\alpha$ are some constants.
We are studying continuum limit of this model. One obtain it by setting size of matrices to infinity and simultaneously tuning the parameter of the action.
  As it was noticed in \cite{Douglas:1989dd},
the partition function of the matrix model with $q-1$ matrices at its $p$-critical point can be described as a solution of higher KdV equation with special initial conditions called the Douglas string equation. Define differential operators
\begin{equation}
Q = d^q + \sum_{\alpha = 1}^{q -1} u_{\alpha}(x)d^{q-\alpha -1},
\end{equation}
\begin{equation}\label{p def}
P  = \left(d^{\frac p q} + \sum_{m=1}^{q-1}\sum_{n= 1 }^{p-1}\tau_{m,n} Q^{ \frac{|pm-qn|}q-1 } \right )_+,
\end{equation}
where $u_\alpha(x)$ are functions in $x$, $d$ is derivative with respect to $x$, differential operators $Q,P$ are polynomials in $d$ and notation $(\dots)_+$ means non-negative powers part of the series expansion in $d$. The constant factors $\tau_{m,n}$ are called times. Note, that terms with $|pm-qn| <q$ don't give any contribution to $P$.  Also two times $\tau_{m,n}$ and $\tau_{q-m,p-n}$ have the same power of Q in (\ref{p def}). Therefore, they contribute in $P$ only in combination  $\tau_{m,n}+\tau_{q-m,p-n}$. We will identify $\tau_{m,n}$ and $\tau_{q-m,p-n}$ in the following. Also we will think that each $\tau_{m,n}$ (with a glance reflection symmetry $\tau_{m,n}=\tau_{q-m,p-n}$) gives contribution in $P$ only once. The Douglas string equation is
\begin{equation}\label{string equation}
 [P,Q] = 1.
 \end{equation}
This is equivalent to the differential equations for the functions $u_\alpha(x)$, from which they could be in principle determined.  The partition function could be found by integrating the equation
\begin{equation}\label{Z equation}
\frac{\partial ^2 Z}{\partial x^2} = u^*_1(x),
\end{equation}
where $u^*_\alpha(x)$ is a suitably chosen solution of (\ref{string equation}). These equations implicitly defines $Z$ as function\footnote{The variable $x$ is one of the times $\tau_{m,n}$ with $|pm-qn| = 1$. These time doesn't contribute in $P$, but it does contribute in $S[u_\alpha]$.  Therefore, it is a constant of integration coming from the integration of $1$ in the right-hand side of (\ref{string equation}).} of the times $\tau_{m,n}$.
  We will use a reformulation of these equations in terms of the action principle introduced in \cite{Ginsparg:1990zc}.  Define an action
\begin{equation}\label{action definition}
 S[u_\alpha] = \text{Res}\, \left(Q^{\frac p q + 1 }+ \sum_{m= 1 }^{q-1} \sum_{n=1}^{p-1}\tau_{m,n}Q^{\frac{|pm-qn|}q} \right),
\end{equation}
where the residue gives the coefficient of $d^{-1}$ taken with a minus sign and integrated over $x$.   The Douglas sting equation reads
\begin{equation}\label{string var eq}
\frac{\delta S[u_\alpha]}{\delta u_\alpha(x)} = 0,
\end{equation}
where $\frac{\delta}{\delta u_\alpha(x)}$ is a variation with respect to $u_\alpha(x)$. The equations (\ref{action definition}) and (\ref{string var eq}) can be found directly by integrating (\ref{string equation}). Additional times with $(m,n)$ satisfying $|pm-qn|<q$ whose don't appear in (\ref{p def}) are the  constants of the integration.

The equations (\ref{Z equation}-\ref{string var eq}) describe a matrix model with $q-1$ matrices near its $p$-critical point. We will refer to it as the $(q,p)$ matrix model. Namely, the critical point correspond to all $\tau_{m,n}$ equal 0 except $\tau_{1,1}=\mu$. The non-zero values of $\tau_{m,n}$ correspond to the deviation from this point\footnote{Times $\tau_{m,n}$ can be understood as coordinates on the space of parameters of the theory. Moreover resonance relations (\ref{res rel intro}) can be considered as non-linear coordinate transformation on this space.} and the KdV time evolutions correspond to flows along the renormalization group trajectories. Thereby, a natural definition of the correlation numbers of the matrix models at the critical point is
\begin{equation} \label{matrix model correlators}
\langle O_{m_1n_1} O_{m_2n_2} \dots O_{m_Nn_N} \rangle = \frac{\partial^N Z}{\partial \tau_{m_1n_1} \dots \partial \tau_{m_{N} n_{N}}}\Big |_{\tau_{1,2}=\tau_{2,1}=\dots = \tau_{q-1,p-1} = 0},
\end{equation}
where the derivative of the partition function are taken at the point where all $\tau_{m,n}$ except $\tau_{1,1}= \mu\ne0$ are zero.
\subsection{Torus partition function}
 The partition function of the matrix models correspond to the summation over all discretized surfaces. It consists of the contribution of the surfaces with fixed genus h\begin{equation}
Z[\tau_{m,n}] = \sum_{h=0}^{\infty} Z_h[\tau_{m,n}].
 \end{equation}
In this subsection we use notation $Z[\tau_{m,n}]$ to underline the partition function $Z$ dependence on times $\tau_{m,n}$ with $m=1,\dots, q$ and $n=1, \dots, p$. In this paper we are interested in the torus ($h=1$) case. In this sense, we need to work out a way to extract its contribution from the all-genus partition function.

 For this purpose we will use the following scaling properties of the partition function \cite{Douglas:1989ve,Ginsparg:1993is,DiFrancesco:1993nw}
\begin{equation}\label{MM partition function}
Z[\varepsilon^{-\frac {2\delta_{m,n}}\gamma}\tau_{m,n}] = \sum_{h=0}^{\infty}\varepsilon^{2(h-1)} Z_h[\tau_{m,n}],
\end{equation}
where $\gamma = 1 +\frac p q$, and gravitational dimensions equal
 \begin{equation}\label{graviational dimensions}
 \delta_{m,n} = \frac{p+q-|pm-qn|}{2q}.
 \end{equation}

 So, the torus partition function is the $\varepsilon^0$ term in the series expansion of   $Z[\varepsilon^{-\frac {2\delta_{m,n}}\gamma}\tau_{m,n}]$ in the variable $\varepsilon$. In the string equations (\ref{Z equation}) and (\ref{string var eq}) one must multiply the times by the $\varepsilon$ factor in the definition of (\ref{action definition}).
  Also, the factors $\varepsilon^{-\frac {2\delta_{m,n}}\gamma}$ in the string equations can be obtained by the formal replacement $\frac d {dx} \rightarrow \varepsilon \frac d {dx}$ in the definition of $Q$. It could be proved with the use of the dimensional analysis.

We are to calculate the partition function for the $(3,p)$ matrix model in the following. One should remember that $q=3$ in this case.  Using the arguments above one can find the genus one partition function as the order $\varepsilon^0$ of $Z$ which solves the string equation with the replacement $\frac d {dx} \rightarrow \varepsilon \frac d {dx}$. For us it will be convenient to indicate the dependence on $\varepsilon$ as a new variable in the functions. For instance, $S[u_\alpha,\varepsilon\,| \tau_{m,n}] = S[u_\alpha\,|\varepsilon^{-\frac {2\delta_{m,n}}\gamma}\tau_{m,n} ]$, where we explicitly wrote the $\tau_{m,n}$  dependence of $S[u_\alpha]$. We will also write the $\varepsilon $-dependence of the solution of the string equations $u^*_\alpha(x,\varepsilon)$. Since we are interested in the order $\varepsilon^0$ in the partition function, we can keep only terms up to the second\footnote{Expansion of $Z$ starts from the term $\varepsilon^{-2}$, but in the string equation (\ref{Z equation}) there is a factor $\varepsilon^2$ in the left-hand side coming from $x$ differentiation. Therefore, $\varepsilon^0$ term in the partition function correspond to the $\varepsilon^2$ term in the $u_1^*(x,\varepsilon)$.} power of $\varepsilon$ in these functions
 \begin{align}
S[u_\alpha,\varepsilon]&=S^{(0)} [u_\alpha]+\varepsilon S^{(1)} [u_\alpha]+\varepsilon^2S^{(2)} [u_\alpha ]+\dots,\\
u^*_\alpha(x,\varepsilon)& =   { u^*_\alpha}^{(0)}(x)+ \varepsilon{ u^*_\alpha}^{(1)}(x)+ \varepsilon^2{ u^*_\alpha}^{(2)}(x)+\dots.
 \end{align}

The evaluation of the genus one partition function $Z_1$ is technical and lengthy. It could be found in the appendix \ref{partition function calculation}. Here we will only formulate the main points of it.

Firstly we want to find the action $S[u_\alpha, \varepsilon]$ up to the second order in $\varepsilon$. To do it, one needs an expression for $\text{res\,} Q^{\frac k 3}$ for any $k>0$. To find it we derive the recurrence equations which relates $\text{res\,} Q^{\frac k 3}$ and $\text{res\,} Q^{\frac k 3+1}$. Solving these equations we can find the action $S[u_\alpha,\varepsilon]$ up to the second order in $\varepsilon$. Next, we solve the string equation up to the second order in $\varepsilon$. Note that the high-order terms ${ u^*_\alpha}^{(1)}$ and ${ u^*_\alpha}^{(2)}$ are expressed in the terms of $S^{(0)} [u_\alpha]$, ${ u^*_\alpha}^{(0)}$ and their partial derivatives. The last step is to integrate the equation (\ref{Z equation}) with ${ u^*_\alpha}^{(2)}$ in the right-hand side. The result is
\begin{equation} \label{3p ZZ}
Z_1^{(3,p)} = -\frac 1 8 \log \det \frac{ \delta^2S^{(0)}[u_\alpha]}{\delta u_\alpha\delta u_\beta}  \Big |_{u_\alpha = {u^*_\alpha}^{(0)}}.
\end{equation}
where $\alpha,\beta$ take values $ 1,2$, the determinant is computed with respect to $\alpha,\beta$ and the variations are taken at the point $u_\alpha(x)= {u_\alpha(x)^*}^{(0)} $.  This is one of the main results of our paper. Note, that since any action of $\varepsilon d$ produce term of the order~$\varepsilon$. Therefore, $S^{(0)}[u_\alpha]$ can be computed directly by taking the residues of $Q$, where $d$ doesn't act on function. It is similar to the calculation of the residues of the ordinary function.

For the $ (2,p)$ models the partition function was obtained in \cite{Belavin:2010pj}
\begin{equation}
Z_{1}^{(2,p)} = -\frac 1 {12}\log \frac{ \delta^2S^{(0)}[u_1]}{(\delta u_1)^2}  \Big |_{u_1 = {u^*_1}^{(0)}}  ,
\end{equation}
These two formulas gives a natural conjecture for the general model torus partition function
\begin{equation}
Z^{(q,p)}_{1}  = - \frac q {24} \log \det  \frac{ \delta^2S^{(0)}}{\delta u_\alpha\delta u_\beta} \Big |_{u = {u^*_\alpha}^{(0)}},
\end{equation}
where $\alpha$ take the values $1,...,q-1$. This conjecture differs by factor $q$ from one made in \cite{Dijkgraaf:1990nc}.

\section{Minimal Liouville Gravity}
 It's known, that any two-dimensional conformal field theory can be coupled to gravity. It was shown by Polyakov \cite{Polyakov:1981rd}, that after choosing conformal gauge for the metric tensor $g_{\mu\nu}= e^{\varphi} \hat g_{\mu\nu}$, a full theory splits in two almost independent parts: the initial CFT part, and the Liouville theory of the scalar field $\varphi$ with action
 \begin{equation}
 S_L = \frac 1 {4\pi} \int_M \sqrt{\hat g}\left(\hat g^{\mu\nu} \partial_\mu \varphi \partial_\nu \varphi + Q\hat R \varphi+ 4\pi \mu e^{2b\varphi}\right)d^2x,
 \end{equation}
 where $\mu$ is a cosmological constant, $M$ is a two-dimensional manifold, $\hat R$ is a Ricci scalar computed for $\hat g_{\mu\nu}$, the constants $Q$ and $b$ are connected with the central charge $c_L$ of Liouville theory through the  relation
\begin{equation}
 c_L = 1+6Q^2 = 1+6(b+b^{-1})^2.
 \end{equation}
The central charge of the Liouville theory is related to the central charge of the matter sector throw the anomaly cancellation condition $c_L+c_M = 26$, where $c_M$ is the central charge of the matter sector.

  For this reason, these theories are called Liouville gravity. If the initial CFT is a minimal model of CFT \cite{Belavin:1984vu} then these models called the Minimal Liouville gravity (MLG). We refers the reader for the additional information about the MLG to the other sources\cite{Zamolodchikov:2005fy,Ginsparg:1993is}.

The $(q,p)$ minimal model of CFT \cite{Belavin:1984vu} has the central charge $c_M = 1-\frac{(p-q)^2}{6pq} $. It has a finite number of primary fields $\Phi_{m,n}$ where $m = 1,\dots, q-1$ and $n=1,\dots , p-1$. Because ofthe reflection symmetry $(\Phi_{q-m,p-n} = \Phi_{m,n})$, only half of these fields are independent.  In the MLG these fields become 'dressed'
\begin{equation}
O_{m,n} = \int \Phi_{m,n} e^{2 b \delta_{m,n} \varphi(x)} \sqrt{\hat g} d^2x,
\end{equation}
where $b = \sqrt{q/p}$, the integration is over 2-dimensional manifold and the gravitational dimensions $\delta_{m,n}$ are the same  as in (\ref{graviational dimensions}). This is the only place where the Liouville scalar couples to the matter fields.

In this paper we are interested in the generating partition function
\begin{equation}
Z[\lambda] =  \left \langle \exp \left(\sum_{(m,n)}\lambda_{m,n}O_{m,n}\right)\right\rangle,
\end{equation}
where $\langle \dots \rangle $ means the functional integral with action $S= S_L+S_M$.
We can think about it  not as the correlation function of the fields, but as the  partition function of the initial theory, whose  action is deformed by the operators $O_{m,n}$ with the couplings $\lambda_{m,n}$.
The correlation numbers of the fields $O_{m,n}$ obeys

\begin{equation}\label{mlg correlators}
\langle O_{m_1n_1} O_{m_2n_2} \dots O_{m_Nn_N} \rangle = \frac{\partial^N Z[\lambda]}{\partial \lambda_{m_1n_1} \dots \partial \lambda_{m_{N} n_{N}}}\Big |_{\lambda_{m,n} = 0},
\end{equation}
where all derivative are taken at the point $\lambda_{m,n}=0$ for all $m=1,\dots, q$ and $p=1,\dots ,n$.

Both the discrete and the continuous approaches are believed to provide the same answers to the similar questions, but as it was noticed in \cite{Moore:1991ir}, there are some arbitrariness because of the contact terms in the correlation functions. This obstacle can be resolve by the use of the non-linear resonance transformation of the form
\begin{equation} \label{resonance relation}
\tau_{m,n} = \lambda_{m,n} + \sum_{m_1,n_1} C^{(m_1,n_1)(m_2,n_2)}_{m,n} \lambda_{m_1, n_1}\lambda_{m_2, n_2} +\dots,
\end{equation}
where all terms satisfy the resonance condition: sum of the gravitational dimension in all  terms must coincide.

In the case $q=3$, the set of primary field $\Phi_{m,n}$ consists of two "rows" with $m=1$ and $m=2$. Due to the reflection symmetry only half of these fields are independent. We will consider first "row" as the independent variables. Also, we will label the fields, the times and the coupling constants as $\Phi_{k}= \Phi_{1,k+1}$, $O_k = O_{1,k+1}$, $\tau_k = \tau_{1,k+1 }$ and $\lambda_k = \lambda_{1,k+1}$  where $k=0,\dots,p-2$.

\section{Correlation Numbers Calculation}
In the following, we are to the calculate correlation numbers of $(3,p)$ models in both the matrix models and in the MLG. This is another result of our paper. We will use notations $u(x)= u_1(x)$ and $v(x)=u_2(x)$ in the following.
\subsection{Matrix Models Correlation Numbers}
In the matrix models one should use the relation (\ref{matrix model correlators}) for the computation of the correlation numbers. The calculation is straightforward, so we will present only results. One must take the derivatives of (\ref{MM partition function}) with respect to the times $\tau_{m,n}$. There are two tricky points in it. The first point is the root of the string equations. As in \cite{Belavin:2013nba} we will choose one special root. Namely, the root with ${v^*}^{(0)}=0$ when $\tau_{m,n}=0$ for $(m,n)\ne (1,1)$. The second point is that one should remember that the root also depends on the times $\tau_{m,n}$.

  All the following formulas are given for $k_i<s$, where $s$ is the quotient of p divided by 3.
We will use normalized $S^{(0)}[u_\alpha]$ and $\tau_k$ in the way that $\frac{\delta S^{(0)}}{\delta u}\Big |_{v = 0}$ and $\frac{\delta S^{(0)}}{\delta v}\Big |_{v = 0}$ will have  coefficients 1 in all its terms. Also, it is convenient to rewrite (\ref{3p ZZ}) using the relation $\frac{\delta^{2} S^{(0)}}{\delta u^2} = \frac {u} 3 \frac{\delta^{2} S^{(0)}}{\delta v^2}$ proven in \cite{Belavin:2013nba}. The partition function then reads
\begin{equation}\label{Z 3p}
Z = - \frac 1 8 \log (\frac {3} u {S_{uu}^{(0)}}^2 - {S_{uv}^{(0)}}^2),
\end{equation}
where we used the subscripts for the partial derivatives. We will also omit  the superscript $(0)$ in following and we will use the notation $S_k$ for $\frac{\partial S}{\partial \tau_k}$. For example, $S_{uvk}$ is $\frac{\partial}{ \partial \tau_k}\frac{ \delta^2 S^{(0)}}{ \delta u \delta v}$. The partition function has the properties
\begin{equation}
S_v(u,0) = 0 \qquad \text{for $p$ - even},
\end{equation}
\begin{equation}
S_u(u,0) = 0 \qquad \text{for $p$ - odd}.
\end{equation}
More generally, the string action has the following property under the reflection of $v$
\begin{equation} \label{action parity}
S_{k_1k_2\dots k_n}(u,-v) = (-1)^{p+\sum k_i} S(u,v)
\end{equation}

We also need the derivatives of the sting equation solution $(u^*,v^*)$ with respect to times at the point $\tau=0 $. It could be  derived by taking the derivatives of the sting equation (\ref{string equation}) and use of (\ref{action parity})

\begin{center}
\begin{tabular}{|c|c|c|}
\hline
 & $p$ - even & $p$ - odd\\
\hline
 $ k $ - even & $\partial_k u^* = - \frac {S_{uk}} {S_{uu}} $& $\partial_k u^* = - \frac {S_{vk}} {S_{uv}}$ \\
\hline
 $ k$ - odd & $\partial_k v^* = - \frac {S_{vk}} {S_{vv}} $& $\partial_k v^* = - \frac {S_{uk}} {S_{uv}}$ \\
\hline
\end{tabular}
\end{center}
where all the functions are taken in the point $(u^*,v^*)$, $\partial_k$ is $\frac \partial {\partial \tau_k}$ and all other first-order derivatives are zero.
\subsubsection{One-point Correlation Numbers}
Let's first consider the case of $p$ and $k$ even.
Taking derivatives of (\ref{3p ZZ}) with respect to $\tau_k$, we derive
\begin{equation}\label{onepoint}
\begin{aligned}
\partial_k Z = -\frac 1 8 \frac 1 {\frac {3} {u^*} S^2_{uu} - S_{uv}^2}\left(\frac 6{u^*} S_{uu} S_{uuk}+\frac 6{u^*} S_{uu} S_{uuu}\partial_k u^*+\frac 6{u^*} S_{uu} S_{uuv}\partial_k v^*\right.\\ \left. -\frac 3{{u^*}^2} S_{uu} S_{uu} \partial_k u^*
-2 S_{uv} S_{uvk}-2 S_{uv} S_{uv}S_{uuv}\partial_ku^*-2 S_{uv} S_{uv}S_{uvv}\partial_kv^*\right),\end{aligned}
\end{equation}
 This expression can be simplified with use of (\ref{action parity}) and the expressions for the partial derivatives of $u^*$ and $v^*$. In the case of $p,k$ even, we have $S_{uv} = 0$, $S_{uvk} = 0$ and $\partial_k v^* = 0$. The formula (\ref{onepoint}) then reads
\begin{equation} \label{p ev k ev}
\partial_k Z = -\frac 1 4 \left( \frac {S_{uuk}}{S_{uu}} - \frac{S_{uuu}S_{uuk}}{S_{uu}^2} +\frac{1}{2u^*} \frac{S_{uk}}{S_{uu}}\right)
\end{equation}

Using this we can find the one-point correlation number  for $k<s$ and $p,k$ -- even
\begin{equation}
\langle O_k\rangle = \frac 1 {24} (p+ 3k-1 )(-\mu)^{- k/2 -1}.
\end{equation}
For $p$ -- even and $k$ -- odd, the one-point function is zero, since the derivatives of $S_{uu}$, $S_{uv}$ and $\partial_k u$ are zero.

In the case of $p$ -- odd we can use the formula (\ref{onepoint}) . But in this case the other terms contribute. With use of $S_{uu} =0$, $\partial_k u^*=0$ the expression for the one-point function in the case of $k$ -- even reads
\begin{equation}\label{1point p odd}
\partial_k Z = -\frac 1 4 \left( \frac {S_{uvk}}{S_{uv}} - \frac{S_{uuv}S_{uvk}}{S_{uv}^2}\right)
\end{equation}
After using the exact expression for $S$ we have in case of $k$ -- even and $k<s$
\begin{equation}
 \langle O_k\rangle = \frac 1 {24} (p+ 3k-1 )(-\mu)^{- k/2 -1}
 \end{equation}
Again it could be seen from (\ref{1point p odd}) that the one-point function $\langle O_k \rangle$ for $k$ -- odd is zero.

Summarising the results of this section we have
\begin{center}
\begin{tabular}{|c|c|c|}
\hline
 & $p$ - even & $p$ - odd\\
\hline
 $ k $ - even & $\langle O_k\rangle = \frac 1 {24} (p+ 3k-1 )(-\mu)^{- k/2 -1} $& $\langle O_k\rangle = \frac 1 {24} (p+ 3k-1 )(-\mu)^{- k/2 -1}$ \\
\hline
 $ k$ - odd & $ \langle O_k\rangle=0$& $\langle O_k\rangle=0$ \\
\hline
\end{tabular}
\end{center}
\subsubsection{Two-point Correlation Numbers}
In this section we don't want to write the full expression for the partial derivative of $Z$ as we did in (\ref{onepoint}). Instead, we will note that the two-point function can be obtained by taking the derivatives of a bit more simple expression. Namely, in the case of $p, k_i$ -- even or $p, k_i$ -- odd
\begin{equation}
\partial_{k_1}\partial_{k_2}Z =-\frac 18  \partial_{k_1}\partial_{k_2} \ln \frac 3 u^{*} S_{uu}^2.
 \end{equation}
 By simple calculation one can find that all the terms coming from $S_{uv}^2$ in (\ref{Z 3p}) doesn't contribute.  On the other hand, in case of $p$ -- even and $k_i$ -- odd or  $p$ -- odd and $k_i$ -- even, only term coming from $S_{uv}^2$ in (\ref{Z 3p}) contribute. Therefore, we can use
 \begin{equation}
 \partial_{k_1}\partial_{k_2}Z =-\frac 14  \partial_{k_1}\partial_{k_2} \ln  S_{uv}
 \end{equation}
 in this case. And finally, all the correlation functions with $k_1$ -- odd and  $k_2$ -- even are zero.

For the two-point correlation function we have
\begin{equation}	
\langle O_{k_1}O_{k_2} \rangle = \frac 1 {48} \left((5+p)(k_1+k_2)+3(k_1^2+k_2^2) +3k_1k_2 +2(p-1) \right)(-\mu)^{(4+k_1+k_2)/2}
\end{equation}
for $p,k_i$ -- even and $k_i<s$.
\begin{equation}
\langle O_{k_1} O_{k_2}\rangle = -\frac 1 {432} (-1+3k_1 +p)(-1+3k_2 +p) (-\mu)^{-(k_1+k_2)/2-2}
\end{equation}
for $p$ -- even,$k_i$ -- odd and $k_i<s$.

\begin{equation}
\langle O_{k_1}O_{k_2} \rangle = \frac 1 {48} \left((5+p)(k_1+k_2)+3(k_1^2+k_2^2) +3k_1k_2 +2(p-1) \right)(-\mu)^{(4+k_1+k_2)/2}
\end{equation}
for $p$ -- odd, $k_i$ -- even and $k_i<s$. And
\begin{equation}
\langle O_{k_1} O_{k_2}\rangle = -\frac 1 {48} (-2+3k_1 +p)(-2+3k_2 +p) (-\mu)^{-(k_1+k_2)/2-2}
\end{equation}
for $p$ -- odd, $k_i$ -- odd and $k_i<s$.

\subsection{Liouville Correlation Numbers}
Lets proceed to the MLG correlation numbers. The main issue in the calculations of the MLG correlators with use of the matrix models results is to find all coefficients in the relation (\ref{resonance relation}) between $t_{m,n}$ and $\lambda_{m,n}$. The criteria for this is the fusion rules of the MLG which results in the equation for the coefficients in (\ref{resonance relation}). Some first of them were determined in \cite{Belavin:2013nba} and we will use their results in the following.

More precisely, we will use the following expressions \cite{Belavin:2013nba} for partial derivatives of action $S$ with respect to couplings $\lambda$:
\begin{equation}\label{S_k ev}
\frac \partial {\partial \lambda_k} S_u = x^{2(p_0-1)} P^{(0,\frac 2 3 (2p_0-3))}_{\frac{s-k-p_0}2}(y) \quad \text{for $(p+k)$--even $k<s$}
\end{equation}
\begin{equation}\label{S_k od}
\frac \partial {\partial \lambda_k} S_u = x^{2-p_0)} P^{(0,\frac 2 3 (1-p_0))}_{\frac{s-k-p_0-3}2}(y) \quad \text{for $(p+k)$--odd $k<s$}
\end{equation}
\begin{equation}\label{S_kk}
\frac {\partial^2} {\partial \lambda_{k_1}\partial \lambda_{k_2}} S_u=\frac 1 p \sum_{k=0}^{\frac{s-k_1-k_2-p_0-2} 2}(6k+4p_0-3)x^{2(p_0-1)} P^{(0,\frac 2 3(2p_0-3))}_k(y)
\end{equation}
The last formula is for $k_i<s$ and $k_i,p$ -- even. We used the notations $P^{a,b}_n(y)$ for the Jacobi polynomials, $x= \frac u u_0$, $y=2x^3-1$, the derivatives of $S$ are calculated at the point $(u,v)= (u^*,v^*)$ and $s$, $p_0$ are the divisor and the remainder of $p$ divided by 3, i.e. $p=3s+p_0$.

The computations in the Liouville frame is quite the same to the matrix model ones. The main difference is that derivatives should be made with respect to $\lambda_{m,n}$.  Therefore, we are to use the same formulas (\ref{p ev k ev}), (\ref{1point p odd}), but with partial derivatives of $S$ taken from (\ref{S_k ev}),(\ref{S_k od}). We will rescale $\lambda_{1,1}= \mu=1$ for simplicity.

The results for the zero and the one-point correlation numbers are
\begin{equation}
\langle 1 \rangle = - \frac 1 4 \log p,
\end{equation}
\begin{equation}
\langle O_k\rangle = - \frac {2+6k+3k^2 - 2(k+1)p} {16p},
\end{equation}
where $k,p$ -- even numbers. The  one-point function for $k$ -- odd is zero.

The two-point correlation numbers are
\begin{align}
\begin{split}
\langle O_{k_1} O_{k_2}\rangle = -\frac 1 {32p^2} \left(9 k_1^3 (1 + k_2) + 9 k_1^2 (1 + k_2) (4 + k_2) +
 3 (2 + k_2) (2 + 3 k_2 (2 + k_2))\right. \\ \left. +
 3 k_1 (1 + k_2) (14 + 3 k_2 (4 + k_2)) -
 6 (1 + k_1) (1 + k_2) (2 + k_1 + k_2) p\right),
\end{split}
\end{align}
where $k_1,k_2,p$ -- even numbers.

All these results depend on the normalisation of the fields and the partition function. Since the relative normalisation of the fields in our paper is the same with ones in \cite{Belavin:2013nba}, we can introduce the normalisation independent quantities
\begin{equation}
\sqrt{\frac{Z^{\text{Sphere}}_{0}}{Z^{\text{Sphere}}_{kk}}} \frac{Z^{\text{Torus}}_{k}}{Z_0^{\text{Torus}}} = \sqrt{\frac{p(p-3(k+1))}{(p+3)(p-3)}}\,\frac {2+6k+3k^2 - 2(k+1)p} {4p \log p}
\end{equation}
\begin{equation}
\begin{split}
\frac{Z^{\text{Sphere}}_{0}}{\sqrt{Z^{\text{Sphere}}_{k_1k_1}Z^{\text{Sphere}}_{k_2k_2}}} \frac{Z^{\text{Torus}}_{k_1k_2}}{Z_0^{\text{Torus}}} =
 \sqrt{\frac{p^2(p-3(k_1+1))(p-3(k_2+1))}{(p+3)^2(p-3)^2}} \\ \frac 1 {8p^2 \log p} \left(9 k_1^3 (1 + k_2) + 9 k_1^2 (1 + k_2) (4 + k_2) +
 3 (2 + k_2) (2 + 3 k_2 (2 + k_2))\right. \\ \left.+
 3 k_1 (1 + k_2) (14 + 3 k_2 (4 + k_2)) -
 6 (1 + k_1) (1 + k_2) (2 + k_1 + k_2) p\right),
\end{split}
\end{equation}

\section{Conclusion}
In the present paper we have calculated the partition function of the $(3,p)$ matrix models and have used it to compute the one- and two-point correlation numbers in the matrix models and the minimal Liouville gravity. We underline that it is not yet known how to calculate correlation numbers in the MLG directly and our paper presents an indirect methods to determine them. We argue that this method can be generalized to the case of arbitrary $(q,p)$. For example, one can use the recent results of \cite{Belavin:2014cua,Belavin:2014hsa} to find the torus correlation numbers of the unitary models.

{\bf{Acknowledgements.}} Author is grateful to A. Belavin and I. Polyubin for useful comments and discussions. I also thanks M. Lahskevich and Ya. Pugai for their comments on the draft of this paper.  The research was performed under a grant funded by Russian Science Foundation (project
No. 14-12-01383).

\appendix

\section{Appendix A: Approximate solution of string equation}\label{partition function calculation}
In this appendix we are to find solution of the string equation up to the second order in $\varepsilon$. We will consider only the  $q=3$ case, and we will use the notations $u(x)=u_1(x)$, $v(x)=u_2(x)$. I

The action is defined as
\begin{equation}
 S[u_\alpha,\varepsilon] = \text{Res}\, \left(Q^{\frac p 3 + 1 }+ \sum_{m= 1 }^{2} \sum_{n=1}^{p-1}\tau_{m,n}Q^{\frac{|pm-3n|}3} \right),
\end{equation}
where
\begin{equation}
Q = \varepsilon^3 d^3 + \varepsilon  u(x) d + v(x).
\end{equation}

To find the solution of the sting equation up to the second order in $\varepsilon$ we need to find, the action up to the same order
\begin{equation}
S[u_\alpha,\varepsilon]= S^{(0)}[u_\alpha] +\varepsilon S^{(1)}[u_\alpha]+\varepsilon^2 S^{(2)}[u_\alpha]+\dots.
\end{equation}
Therefore, we need to learn how to find the residues of $Q^{l + \alpha/3}$ for an arbitrary values of $l$ and $\alpha$.
 One can do it with use of the following technique \cite{Ginsparg:1993is,DiFrancesco:1993nw}. The idea is to find the recurrence relation for the residues of this form.

We parametrise the negative power part as $Q^{l+\alpha/3}_- = \{R_l ,d^{-1}\}+\{K_l ,d^{-2}\}+\{M_l ,d^{-3}\}+\dots$. We have used the notation $\{a,b\} = ab+ba$. Since, any operator commutes with any its power and also using relation $Q^{l+\alpha/3}=Q^{l+\alpha/3}_-+Q^{l+\alpha/3}_+$, where $Q^{l+\alpha/3}_+$ is the non-negative power part in $d$, we have
\begin{equation}
[Q^{l+\alpha/3}_+,Q]= [Q, Q_-^{l+\alpha/3}].
\end{equation}
The left-hand side contains only non-negative powers of $d$, while the right-hand side contains only the finite number of non-negative powers of $d$. They are
\begin{equation}\label{comm q q}
[Q^{l+\alpha/3}_+,Q]=[Q, Q_-^{l+\alpha/3}] = 6 \varepsilon R'_l d + 6\varepsilon K_l'+3\varepsilon^2R''_l.
\end{equation}
Also we can find that
\begin{equation}
Q^{l+1+\alpha/3}_+ = (Q^{l+ \alpha/3} Q)_+= Q^{l-2/3} Q + 2 R_l \varepsilon^2d^2 + (2K_l-\varepsilon R'_l)\varepsilon d + 2M_l+\varepsilon^2R''_l-2\varepsilon K_l'+2R_l u.
\end{equation}
Commutating both side of this equation with $Q$ and using (\ref{comm q q}), we obtain three recurrence equations on three unknown functions $K_l$, $M_l$ and $R_l$. One of them can be trivially solved to give the expression of $M_l$ in the terms of $R_l$ and $K_l$. Other two reads

\begin{align}\label{rec eq}
\begin{split}
6\varepsilon R_{l+1}' =4\varepsilon ^3 K'''_l+2 \varepsilon u' K_l+4 \varepsilon u K'_l-2\varepsilon ^2 u''R_l+4\varepsilon v'R_l-3\varepsilon^2 u'R'_l+6\varepsilon v R' _l\\
6\varepsilon K_{l+1}'=  2\varepsilon  v' K_l+ 6\varepsilon vK'_l-\frac 1 3 \varepsilon ^3 u'''R_l
-\varepsilon ^2u''K_l-3\varepsilon^2u'K'_l-\frac 3 2 \varepsilon^3 u''R'_l-\frac 5 2\varepsilon^3  u'R''_l\\ -\frac 5 3 \varepsilon^3 uR'''_l-\frac 4 3\varepsilon uu'R_l-\frac 4 3 \varepsilon u^2R'_l-\frac 1 3\varepsilon^5 R^{(5)}_l
\end{split}
\end{align}
Note, that these relation are the same for the different values of $\alpha$ in $\text{Res}\,Q^{l+\alpha/3}$, but they differ in initial conditions. This equation could be solved by using the following ansatz
\begin{align}\label{ansatz 1}
\begin{split}
R_l^{(1/3)}& = \sum_{k\ge 0} (A^{(l)}_k u^{3k+1}v^{l-1-2k} + \varepsilon B^{(l)}_k u^{3k+1} v^{l-2-2k} u' +\varepsilon^2 D ^{(l)}_k u^{3k+2}v^{l-3-2k}u''+\varepsilon^2 E^{(l)}_k u^{3k+1}v^{l-3-2k}u'^2\\&+\varepsilon^2 F ^{(l)}_k u^{3k+2}v^{l-4-2k}u' v'+\varepsilon^2 G^{(l)}_k u^{3k}v^{l-3-2k}v'^2+\varepsilon^2 H ^{(l)}_k u^{3k}v^{l-2-2k}v'') + \dots,
\\
K_l^{(1/3)}&= \sum_{k\ge0}( a^{(l)}_k u^{3k} v^{l-2k}+ \varepsilon b^{(l)}_k u^{3k} v^{l-1-2k} u' +\varepsilon^2 d^{(l)}_k u^{3k+1}v^{l-2-2k}u'' +\varepsilon^2 e^{(l)}_k u^{3k}v^{l-2-2k}u'^2 \\&+\varepsilon^2 f^{(l)}_k u^{3k+1}v^{l-3-2k}u'v' +\varepsilon^2 g^{(l)}_k u^{3k+2}v^{l-4-2k}v'^2 +\varepsilon^2 h^{(l)}_k u^{3k+2}v^{l-3-2k}v'')+\dots,
\end{split}
\end{align}
where dots means terms with the higher $\varepsilon$ powers and we presented ansatz only for the case $\alpha =1$.
Substituting these equations in (\ref{rec eq}), we derive a bunch of
equations for the coefficients in (\ref{ansatz 1}). They can be solved by some cumbersome calculation.  The answer takes the simple form
\begin{equation}
\begin{aligned}
S[u,v] = \text{Res}\, \left(Q^{\frac p 3 + 1 }+ \sum_{m= 1 }^{2} \sum_{n=1}^{p-1}\tau_{m,n}Q^{\frac{|pm-3n|}3} \right)=  S^{(0)}-\frac 1 2 \varepsilon S_v^{(0)} u'-\frac 1 6\varepsilon^2 S_{vv}^{(0)} u u'' \\- \frac 1 {12 }\varepsilon^2 S_{uvv}^{(0)} u u'^2 - \frac 1 6 \varepsilon^2 S_{vvv}^{(0)} uu'v' + \frac 1 4\varepsilon^2 S_{uvv}^{(0)}v'^2 + \frac 1 2 \varepsilon^2 S_{uv}^{(0)}v''+\dots,
\end{aligned}
\end{equation}
where subscripts of $S^{(0)}[u,v]$ mean the variations with respect to the functions $u(x)$ and $v(x)$, dots means the higher terms in $\varepsilon$.
Our next step is to solve the Douglas equation (\ref{string var eq}) which could be represented as ${u^*}(x,\varepsilon) = {u^*}^{(0)}(x) + \varepsilon^2 {u^*}^{(2)}(x) + O(\varepsilon^2)$. The last step is to integrate (\ref{Z equation}) with $u_*$ replaced with ${u^*}^{(2)}$. The result is (\ref{MM partition function}).

\end{document}